\documentclass[aip]{revtex4}
\usepackage{amsmath}
\usepackage{graphicx}
\begin{document}

\title{An upper limit on additional neutrino mass eigenstate in 2 to 100 eV region from ''Troitsk nu-mass'' data}

\author{A.\,I.\,Belesev}
\author{A.\,I.\,Berlev}
\author{  E.\,V.\,Geraskin}
\author{ A.\,A.\,Golubev}
\author{ N.\,A.\,Likhovid}
\author{ A.\,A.\,Nozik}
\author{ V.\,S.\,Pantuev}
\author{ V.\,I.\,Parfenov}
\author{A.\,K.\,Skasyrskaya}
\affiliation{Institute for Nuclear Research of Russian Academy of Sciences, 117312 Moscow, Russia}
%\date{\today}{}

\begin{abstract}
We  performed a search for any sign of an additional neutrino mass state in $\beta$-electron spectrum based on data reanalysis of direct electron antineutrino mass measurements in Tritium beta-decay in the Troitsk nu-mass experiment. The existing data set allows us to search for such a state  in the mass range up to $100~eV$. The lowest value at a 95\% C.L. upper limit for the contribution of a heavy eigenstate into electron neutrino is around or less than 1\% for masses above $20~eV$.
\end{abstract}

%%% PACS numbers

\maketitle
%Introduction
Throughout the last couple decades it has become clear that the Standard Model (SM) of elementary particles cannot explain some of the observed phenomena in particle physics, astrophysics and cosmology. These are baryon asymmetry, dark matter, neutrino oscillations  and others. Neutrino oscillations from short baseline experiments favor the existence of an additional neutrino mass state to the three active neutrinos in the SM. Astrophysical observations and cosmology also point to the fact that the effective number of neutrinos is greater than 3~\cite{shaposhnikov}. This can be interpreted as a possible existence of at least one sterile neutrino.  Sterile neutrino is a natural consequence of the non-zero neutrino mass and appears  in many theories beyond the SM. From this point of view,  addition of a. Would a sterile neutrino be found it will be the first particle beyond the Standard Model. While there is a number of the results which disfavor or even are in contradiction with the hypothesis of sterile neutrino, many experiments are undergoing or are planning to search for them. For details we refer to the "Light sterile neutrinos: a white paper''~\cite{white_paper}.  

It becomes important to check all possible experimental data to  prove,  disprove or set an upper limit for the sterile neutrino hypothesis. Results of the reanalysis of our data on the direct electron antineutrino mass measurements in Tritium $\beta$-decay in the Troitsk nu-mass experiment~\cite{our_nu_e} are presented in this paper. The group led by V. M. Lobashev was obtaining these data in the period of 1997-2004. We  used the same file set and analysis framework as for the electron antineutrino mass. We performed a search for any sign of an additional neutrino mass state in the $\beta$-electron spectrum. Such a state with a finite mass would exhibit itself as a kink in the spectrum. Recently a similar analysis was published
based on the Mainz data~\cite{mainz}.

%Experiment
The Troitsk experiment has two major parts: an integrating electrostatic spectrometer with adiabatic magnetic collimation and   
a windowless gaseous tritium source  as a volume for $\beta$-decays.  The spectrometer resolution was about 4 eV. We measured an integrated electron spectrum at the region of the last 200-300 eV from the spectrum endpoint (18575 eV) by varying the electrostatic potential $V$ on the spectrometer electrode. 
All details on experimental setup, data taking, analysis, corrections and estimation of systematic error are published in Ref.~\cite{our_nu_e}. For electron antineutrino mass squared we published the value $m_{\nu }^{2}=-0.67\pm 2.53\;\mathit{eV}^{2}$.
 
% Data analysis
In accordance with Ref.~\cite{our_nu_e}, the spectrum of electrons in tritium  $\beta$-decay is the following:
%\begin{multline}
\begin{equation}
S(E,E_0,m_\nu^2) = NF(E)(E+m_{e})p_e(E_0-E)\cdot \sqrt{(E_0-E)^2-m_\nu^2},
\label{eq:one}
\end{equation}
where $N$ is the normalization constant, $F(E)$ - the so called Fermi-function responsible for electrostatic interaction between electron and nucleus,  $E$ and  $p_e$ stand for the electron energy and momentum, $E_0$ - for the beta-spectrum endpoint and $m_\nu$ - for the neutrino mass.
After decay of a tritium nucleus the primary molecule of T$_2$ becomes a molecule of T$^3$He. Often, with a probability of about 43\%,  T$^3$He does not go to its ground state, thus we have to sum over all molecule final states $i$ and eq.~\ref{eq:one} 
should be replaced by the sum:
\begin{equation}
S(E,E_0)=\sum_i S(E, E_0-\varepsilon_i)\cdot P_i , 
\label{eq:states}
\end{equation}
where $\varepsilon_i$ is the energy of the excited state and $P_i$ is its probability, the sum of  $P_i$ equals one.  
Finally, we get the following expression for the experimental integrated electron spectrum versus retarding potential on the spectrometer electrode $V$:
\begin{equation}
Sp(V)=N\cdot \int \big[S(E,E_0,m_{\nu}^2)\otimes Tr(E) \big]\cdot R(V,E)dE + bkgr ,
\label{eq:spectrum}
\end{equation}
where $S(E,E_0,m_{\nu}^2)$ is the electron spectrum from the $\beta$-decay, Eq.~\ref{eq:states}; $Tr(E)$ is the energy loss spectrum and $R(V,E)$ is the resolution function (see~\cite{our_nu_e} for details),  and $bkgr$ is the experimental background. 

If the number of neutrino eigenstates is larger than 3, for the effective electron neutrino we can write $\mid\nu_e\rangle=\sum\limits_i  U_{ei}\mid\nu_i\rangle$, where $U_{ei}$  are the mixing matrix elements. We restrict ourselves to one additional heavy neutrino ($i=4$).  From neutrino oscillation results it is known that the mass splitting between active neutrinos is much less than one electronvolt. Thus, masses of "normal" eigenstates are probably negligibly small,  and one can assume $m_1=m_2=m_3=0$. Consequently, the electron spectrum with one additional heavy neutrino component can be written in the following way:
\begin{equation}
S(E)dE = NF(E)(E+m_{e})p_e(E_0-E)\cdot (U^2_{e4}\sqrt{(E_0-E)^2-m_4^2}+(1-U^2_{e4})(E_0-E)),
\end{equation}
where $U^2_{e4}$ is the fraction of the heavy neutrino in the electron neutrino and $m_4$ is the mass of the heavy neutrino eigenstate. In other words, we fit the spectrum with an assumption that its  major component has a relative amplitude $1-U^2_{e4}$ and is attributed to zero neutrino mass, besides there is an additional feature with the relative amplitude $U^2_{e4}$ for heavy mass $m_4$. It is worth mentioning that in a usual notation for the neutrino  oscillations parameter $sin^2(2\theta)$~\cite{oscill}, at small $U^2_{e4}$ there is an approximate relation $sin^2(2\theta) \approx 4 U^2$.

To get an upper limit for $U^2_{e4}$ the Bayesian approach has been used for the parameter estimation. The posterior probability $L$ for parameter $U^2_{e4}$ is calculated as a product of posterior probabilities $L_k$ for different experimental runs ($L = \prod L_k$). For each run the probability calculation procedure is the following: 

\begin{enumerate}
\item At first we set $U^2_{e4}$  to zero and fit three spectrum parameters: $E_0$, $N$ and $bkg$. This is required to get a precise region for the additional parameters. One must note that, while $E_0$ is a physical value and should not change from run to run, in practice it is a free parameter and depends on the spectrometer calibration and can vary for different data sets.
\item Next step: set $m_4^2$ and get a four-dimensional likelihood function $L(U^2_{e4}, E_0, N, bkg).$
\item In order to take into account possible correlations between parameters, we marginalize the likelihood function over all non-essential parameters: $L(U^2_{e4}) = \int\limits_{E_0}\int\limits_{N}\int\limits_{bkg} L(U^2_{e4}, E_0, N, bkg)$. Due to the fact that the calculation of the likelihood function for one set of parameters is greatly time consuming, integration is made by Monte-Carlo procedure. The values of $L(U^2_{e4})$ are saved in the table.
\item Repeat the procedure from step 2 for different values of $m_4^2$.
\end{enumerate}

During all calculations we presume that $0 \leq U^2_{e4} \leq 1$. Finally an upper limit has been found by solving the equation:
\begin{equation}
\frac{\int\limits_0^{limit}L(U^2_{e4})}{\int\limits_0^{1}L(U^2_{e4})} = \alpha,
\label{eq:limit}
\end{equation}
where $\alpha$ is the required confidence level, namely 95~\%.

In the current analysis we used only the data in which the spectrometer electrode potential is higher than $E_{low} = 18400~V$. We also checked that the usage of $E_{low} = 18300~V$ does not change the result.

%error estimation

It is worth stressing that all statistical errors and correlations are already incorporated in the upper limit estimation by Eq. \ref{eq:limit}. As for systematical errors, the shift of the upper limit at 95\% C.L. caused by the change of one or more additional parameters within systematic boundaries proved to be negligibly small. A large error  was expected to arise from uncertainty in the final states spectrum (FSS) of T$^3$He, Eq.~\ref{eq:states}. Using the spectrum from \cite{FSS1} and \cite{FSS2} we get practically the same result. It should be emphasized that there are no experimental data for FSS of T$^3$He molecule and in both references FSS was calculated. The whole analysis procedure was tested on simulated data.

%results
\begin{figure}[htb]
\includegraphics[width = 120 mm]{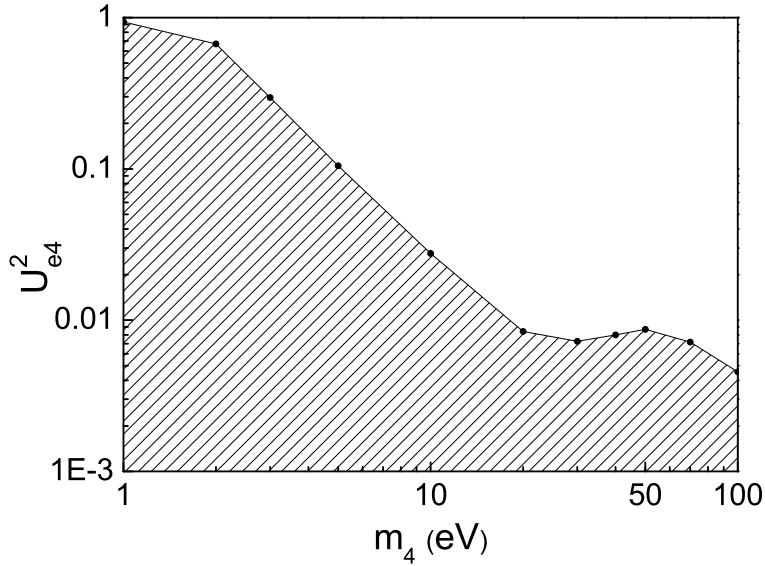}
\caption{The upper 95\% C.L. limit for admixture $U^2_{e4}$ of the heavy neutrino eigenstate in $\beta$-electron spectrum versus its mass, $m_4$.}
\label{fig:limits}
\end{figure}

Results on the upper limit at 95\% C.L. for additional neutrino mass eigenstate, $m_4$, are presented in Fig.~\ref{fig:limits}. As expected, our experiment with a sensitivity limit of about 2 eV~\cite{our_nu_e} has a poor rejection factor at $m_4$ of about a few electronvolts.  At larger masses the limit goes down, then at the range $m_4=$ from 20~eV to 100~eV the upper limit stays between 0.01 and 0.005 .

 The important feature of the raw data processing is the so called bunch rejecting algorithm. Its main purpose is to filter short-timed high intensity "bunch" events which are caused by the electrons trapped in the spectrometer magnetic bottle. The procedure for "bunch" search is automatic and its effectiveness is lower for higher count rates. The count rate in the spectrometer below the spectrum endpoint by around 50 eV is usually critical for bunch rejection algorithm. The simulations show that while bunch rejection parameters cannot affect the estimation of the active neutrino mass, they do affect estimation of the probability for heavy neutrino mass above 30 eV. Our analysis is also not sensitive enough in the region of $m_4$ 50 - 100 eV  because of wider intervals between the measured experimental points (around 25 eV) of electron spectrum moving away from the endpoint. 

In conclusion, we reanalyzed our data of the direct electron antineutrino mass measurements in Tritium $\beta$-decay in the Troitsk experiment.  The file set and the analysis framework were identical to those used in the original work.  The maximum likelihood method was used to evaluate a possible contribution  of the heavy extra mass state  $m_4$ with amplitude $U^2_{e4}$ with the assumption that all three active neutrinos have zero masses. In the mass range $2~<~m_4~<~20~eV$ an upper  limit at 95\% C.L. quickly goes down to $U^2_{e4} = 0.01$ and then stays close to this level  for $m_4$ up to $100~eV$. 

 The current analysis  was supported by RFBR under grant numbers 11-02-00935-a, 12-02-31323-mol-a and 12-02-12140-ofi-m. We also would like to thank  our colleagues V.\,N.~Aseev, N.\,A.~Golubev, O.\,V.~Kazachenko, B.\,M.~Ovchinnikov, N.\,A.~Titov, Yu.\,I.~Zakharov, S.\,V.~Zadorozhny and and I.\,E.~Yarykin for their valuable contribution to experiment preparation and data taking.

\end{document}